# Room-temperature solid state quantum emitters in the telecom range

Yu Zhou[1], Ziyu Wang [1], Abdullah Rasmita[1], Sejeong Kim[2, 3], Amanuel Berhane[2, 3], Zoltán Bodrog[4], Giorgio Adamo[1, 5], Adam Gali[4, 6], Igor Aharonovich [2, 3 *], Wei-bo Gao [1, 5*]

*1 Division of Physics and Applied Physics, School of Physical and Mathematical Sciences, Nanyang Technological University, Singapore 637371, Singapore*

*2 School of Mathematical and Physical Sciences, University of Technology Sydney, Ultimo, NSW, 2007, Australia*

*3. Institute of Biomedical Materials and Devices (IBMD), Faculty of Science, University of Technology Sydney, Ultimo, NSW, 2007, Australia*

*4. Wigner Research Centre for Physics, Institute for Solid State Physics and Optics, Hungarian Academy of Sciences, PO Box 49, H-1525, Budapest, Hungary*

*5. The Photonics Institute and Centre for Disruptive Photonic Technologies, Nanyang Technological University, 637371 Singapore, Singapore*

*6. Dept. of Atomic Physics, Budapest University of Technology and Economics, Budafokiút 8., H-1111, Budapest, Hungary*

**On demand single photon emitters (SPEs) play a key role across a broad range of quantum technologies[1-7], including quantum computation[8], quantum simulation[9-11] quantum metrology[12] and quantum communications[13-15]. In quantum networks and quantum key distribution protocols, where photons are employed as flying qubits, telecom wavelength operation is preferred due to the reduced fibre loss. However, despite the tremendous efforts to develop various triggered SPE platforms[1], a robust source of triggered SPEs operating at room temperature and the telecom wavelength is still missing. Here we report a triggered, optically stable, room temperature solid state SPE operating at telecom wavelengths. The emitters exhibit high photon purity (~ 5% multiphoton events) and a record-high brightness**

**of ~ 1.5 MHz. The emission is attributed to localized defects in a gallium nitride (GaN) crystal. The high performance SPEs embedded in a technologically mature semiconductor are promising for on-chip quantum simulators and practical quantum communication technologies.**

In the last decade, SPEs have been explored as key resources for many quantum technologies, where the photons are employed as "flying qubits". In linear optical quantum computation, the information is carried by single photons that interact with each other through optical elements and projective measurements[8]. In Boson sampling quantum machine, single photons are used as inputs for a series of interferometers to complete photonic circuits[16]. Furthermore, for the majority of quantum key distribution (QKD) systems, single photons are the most fundamental building blocks[15]. However, due to the lack of a practical ultra-bright triggered SPE, current QKD systems widely uses weak coherent pulse (from attenuated lasers). In this regard, a true SPE would still be preferred to achieve a longer secure distance and thus a better performance.

For the above mentioned applications, the operation wavelength is preferred to be in the telecom range due to its lower attenuation loss in fiber transmission performance as compared to shorter wavelengths. Current solutions for triggered SPEs at this wavelength range mostly rely on a variety of semiconductor quantum dots (QDs). For instance, InAs/InP QDs have single photon emission at telecom wavelength[17] and have been utilized to realize 120 km QKD[18], while InAs/InGaAs QDs have been used for the generation of entangled photon pairs[19]. These sources, however, require cryogenic cooling down to liquid helium temperatures, which is not ideal for many scalable and practical devices. In a complementary approach, down conversion of near infrared photons to the telecom range was employed[20]. However, the photon generation is limited by the non-ideal conversion efficiency in the non-linear process. Finally, recent studies of carbon nanotubes (CNTs)[21] showed the potential of hosting room temperature (RT) SPEs in the infrared spectral range, but the emitters suffer from severe problems such as blinking, bleaching and very low photon rate of kHz.

In this work, we overcome all aforementioned deficiencies and report on unprecedented photostable, RT SPEs embedded in a GaN crystal and are operating at the telecom range. The SPEs exhibit both excellent purity, with $g^2(0) \sim 0.05$, and a high brightness exceeding $10^6$ counts/s. A number of GaN defects ranging from the ultraviolet (UV) to infrared (IR) wavelength have been predicted with a density-functional study[22]. However, most of the experimental studies have been focused on defects in the UV[23] and the visible range[24], while emitters in the infrared range remained largely unexplored. Moreover, since nanofabrication procedures with GaN are well established, the discovery of telecom SPEs can be instantly adopted and is highly promising for practical quantum technologies.

Figure 1a shows the GaN crystal structure and an optical image of a typical GaN wafer where SPEs were observed. The studied sample is a 2 μm thick magnesium (Mg)-doped GaN layer on 2 μm undoped GaN layer grown on sapphire. We first study the photoluminescence (PL) spectra of GaN emitters with a home-built confocal scanning setup, where 950 nm diode laser was used to excite the defect and emission above 1000 nm was collected and guided to a spectrometer or a pair of superconducting detectors in a Hanbury Brown-Twiss (HBT) configuration for photon counting (see Methods). Emitters are randomly distributed on the GaN substrate and the confocal PL map around the one of the SPEs, named as SPE1, is shown in figure 1b as an example. This particular defect emits ~ 350 kcounts/s at room temperature. A survey of the sample shows different SPEs with distinct narrowband emission, with zero phonon lines (ZPLs) ranging from 1085 nm to 1340 nm. The distribution of ZPLs will be discussed below. Examples of three SPEs at RT are given at the lower panel of figure 1c, while the top panel shows example of four PL spectra of the SPEs recorded at cryogenic temperature (4K). At RT the full width at half maximum (FWHM) of the emitters' linewidth ranges from 3 nm (ZPL at 1120nm) to 50nm (ZPL at 1285nm) (See Supplementary figure S1 for more information). At 4K, the FWHM of the measured SPEs reduces to a few nm, but is still broadened due to coupling to the lattice phonons.

To confirm the nonclassical photon statistics from the studied emitters, the second order autocorrelation function, $g^{(2)}(\tau)$, is recorded. Figure 1d shows the $g^{(2)}(\tau)$ recorded from SPE1 under continuous wave (cw) laser excitation at RT. The vanishing peak at zero delay time indicates that the emission is non classical and the source is a SPE. Other emitters have similar properties and more data can be found in the supplementary information. The data is fit with a three level model (equation 1)

$$g^2(\tau) = 1 - \alpha * exp(-|\tau|/\tau 1) + \beta * exp(-|\tau|/\tau 2) \tag{1}$$

where $\tau_1$, $\tau_2$ are the excited state and the metastable state lifetimes, respectively, and $\alpha$, $\beta$ are the fitting parameters. The obtained value at zero time delay $g^2(0)$ is $0.05 \pm 0.02$ without any background correction, proving that this is one of the purest RT SPEs. The slight deviation from zero is attributed to the background from other defects within the GaN crystal. Furthermore, as shown by the pulsed $g^{(2)}(\tau)$ in figure 1e, the SPE can be efficiently triggered. To obtain the $g^{(2)}(\tau)$ in a pulsed regime, pulsed laser excitation with 80MHz repetition rate and one ps pulse width was used. The obtained value at zero delay time corresponds to $g^2(0) = 0.14 \pm 0.01$. It is also well below the classical threshold 0.5 for proving SPE. The value of $g^2(0)$ under pulsed excitation is lifted due to higher background in the pulse regime.

To study the performance of the SPEs in more details, a saturation curve is recorded from the emitter labeled SPE1 in figure 1e. More data is provided in the supporting information. Figure 2a shows the saturation behavior of SPE 1 under cw laser excitation. The data are fit well using the three level system with an equation $I(P)=I_\infty \times P/(P+P_s)$, where $I(P)$ is the measured intensity count rate, $P$ is the excitation power; $P_s$ (saturation power) and $I_\infty$ (maximum count) are two fitting parameters. For this particular emitter, we obtain $P_s = 2.32 \pm 0.08$ mW, and $I_\infty = 0.69 \pm 0.01 \times 10^6$ counts per second. Such a count rate is on par with brightest room temperature SPEs from a bulk crystal. The detailed discussion of extraction efficiency analysis is given in the Methods section.

The source photostability is measured under low and near saturation excitation powers. The data is shown in figure 2b over an excitation period of 120 seconds under 0.1 mW (black curve), 0.5 mW (red curve) and 1.8 mW (blue curve) of excitation power, respectively. The time binning in this measurement is 100 ms and no obvious blinking or bleaching has been observed, proving the stability of the SPE. The fluorescence lifetime of the emitter is presented in figure 2c. The measured lifetime is 736 ± 4 ps (black curve), which is in accord with the value obtained from the $g^{(2)}(\tau)$ fitting (776 ± 39 ps). The data was fit with single exponential function and the instrument response function (blue curve) for our setup is shown for comparison in figure 2c. The system behaves according to a three level model (figure 2d) with slight bunching at longer time scales (see supporting information for the detailed rate equations).

To further optimize the brightness of the SPE by means of improving the photon extraction efficiency, we take advantage of the rather mature fabrication techniques of GaN. Rather than employing a top down reactive ion etching techniques, a separate GaN sample is grown on a patterned sapphire substrate (PSS). The scanning electron microscope image of such a structure is shown in figure 3a. This structure is designed and extensively used for the LED emission enhancement by increasing the reflection area and therefore improving the light extraction efficiency.

To study the effect of the PSS on the extraction efficiency, 3D finite difference time domain (FDTD) optical simulation (using Lumerical software) is carried out for the defect embedded in a pristine GaN and the one grown on a PSS. Based on the simulation results, the far field pattern and the collection efficiency can be obtained[25]. For the bare GaN substrate, a model consisting of a GaN layer (refractive index = 2.33) on top of sapphire layer (refractive index = 1.75) in an immersion oil environment (refractive index = 1.52) is assumed. For the GaN/PSS structure, the model is modified to include the patterned geometry (see supplementary figure S3 for an illustration). The FDTD simulation provides the far field radiation patterns for the emitter in a pristine GaN and in a GaN/PSS structure as seen in figure 3b and figure 3c, respectively. Note that the maximum intensity in figure 3c is higher than that in figure 3b. The power collected within $62^0$ half-angle (i.e. the maximum collected half-angle for a 1.35 NA objective in an immersion oil environment) of the far field radiation is then calculated for both cases. By comparing these calculation results, it is found that an enhancement up to ~ 2 times is expected when the PSS is used. This enhancement is attributed to the additional light focusing provided by the V-shaped region between two cones of the PSS.

Upon scanning the sample, similar SPEs emitters with the one shown in figure 1 can be easily detected. This is evident from the confocal map in figure 3d. The blue pattern corresponds to the PSS structure underneath the surface. Figure 3e shows a saturation curve for a SPE on a pristine GaN sample compared to a SPE embedded in the GaN grown on PSS. The increase in count rate is clearly observed. The SPEs located in the center of the PSS pattern exhibits an enhanced count rate reaching $2.33 \times 10^6$ counts/s at saturation, compared to a saturation rate of $1.13 \times 10^6$ counts/s for another SPE in a pristine GaN. The counts for samples on PSS correspond to an enhancement factor of ~ 2 as compared with the ones without PSS, consistent with the Lumerical simulations. Figure 3f shows a statistical count comparison of 10 different emitters that convincingly proves the overall enhancement (details shown in the supplementary information).

Finally, we discuss the possible origin of our emitters. With the wide wavelength distribution, we consider the model of an optically active point defect near cubic inclusions within the hexagonal lattice of GaN[24]. The PL energies are calculated with a simple quasi one-dimension model, where a point defect resides in the proximity of the cubic inclusion. The luminescence is the result of an exciton recombination, where the hole is tightly localized at the point defect while the loosely localized electron's position is more influenced by the potential lineup generated by the stacking sequence of cubic and hexagonal GaN bilayers. Thus, the binding energy of the exciton, and PL wavelength are determined by the relative position of the point defect with respect to the cubic inclusion. Figure 4a, b shows the distribution of the ZPLs obtained for a single and a double cubic inclusion that result in a broad distribution of the ZPLs. The configuration of multiple cubic inclusions fits the presented experimental data well. Interestingly, if the defect is located exactly at the interface of the cubic/hexagonal boundary, PL lines are expected to occur only at the lower (~ 1100 nm) and the higher (~ 1350 nm) wavelength range. Both emission wavelengths were experimentally observed in our experiments (see figure 4c).

While the presented model explains well the ZPL shifts observed in our work, the origin of the luminescent impurity remains unknown. So far, there is no evidence in the literature on a RT narrowband emission (even from ensembles) from GaN in the infrared spectral range. Luminescence at ~ 1.5 µm was reported from GaN wafers implanted with Er ions[26] (the emission is an internal transition of Er), while broad PL (only at cryogenic temperatures) from high energy electron irradiation and annealing was reported at ~ 0.9 eV and was attributed tentatively to Ga interstitials[27]. Other works observed narrowband luminescence around ~1 eV (also at cryogenic temperature) that was attributed to Cr impurities[28, 29]. Further density functional theory (DFT) calculations and targeted ion implantation experiments into an ultra-pure GaN crystals will be required before the absolute crystallographic structure of the quantum emitters can be unveiled.

Overall, we reported on ultra-bright, optically stable, single photon emission in GaN at the telecom range. The SPEs operate at RT and have FWHM as small as ~ 3 nm. Furthermore, we demonstrated that these emitters can be observed in commercially available, pristine GaN wafers without the necessity of electron or ion irradiation, and their emission rates can reach MHz rates upon substrate patterning. Several immediate research directions come out of our work. First, unveiling the origin of the emitters via DFT modeling in conjunction with ion implantation and growth should be feasible and within reach. In addition, doping techniques of GaN are well developed, and optoelectronic components with p-i-n layers are available[30]. It may be possible to realize electrically triggered RT SPE in the telecom range – a highly crucial component for scalable devices[30, 31]. Furthermore, photonic cavities and waveguides can be engineered from GaN relatively easy[32, 33], paving the way to an on-chip integrated quantum nanophotonics platform[4, 34-

[37]. Finally, investigating the spin properties of these defects can become useful for exploring spin-photon interfaces[6] in GaN for quantum information processing.

## Methods

**Sample description**

For the data in figure 1 & 2, we use a GaN with 2 μm thick magnesium (Mg)-doped GaN layer on 2 μm undoped GaN layer grown on planar sapphire. For the data in figure 3, 6.5 μm GaN is grown on a patterned sapphire substrate (PSS) with cone structures. The cone patterns have a width of 2.5 μm and a height of 1.7 μm with a separation distance of 3 μm.

**Experimental setup**

For room temperature measurement, the samples are mounted on a three direction nanopositioner (PI P-611.3S) for confocal PL map scanning. A diode laser with wavelength of 950 nm is used for cw excitation and a Titanium:Sapphire one ps laser with 80MHz repetition rate (Spectra Physics) is used for pulse excitation. The laser beam passes through a 1000 nm short pass filter to clean up the residual spectrum above 1000 nm and then is focused down with an oil objective with N. A. of 1.35 (Nikon). The collected light goes through the same objective into a single mode fiber that is used as a confocal aperture. The signal is then guided to a low jitter (~30 ps) super-conducting-single-photon detectors (Scontel SSPD) for counting or a liquid nitrogen cooled InGaAs camera (Princeton Instruments) for spectrum analysis. For the Hanbury-Brown and Twiss interferometer set-up, the PL signal is divided with a fiber beam splitter and detected by two SSPDs. The correlation coincidence is recorded by a time-correlated photon counting card (Picoharp, PH300). For the cryogenic measurements, the sample is mounted in a closed cycle cryostat (Montana Instruments) equipped with attocube steppers for the rough scanning and galvo mirrors above the cryostat for the fine-scanning.

**Source efficiency and set-up detection efficiency**

For the pristine GaN substrates, based on the FDTD simulation result, we estimate the extraction efficiency of single photon (that is, the fraction of emitted photon collected by the objective) as $\eta_c = 0.13$. Before the detector, there is a transmission loss due to fiber collection (collection efficiency $\eta_f = 0.4$) and objective loss (transmission $\eta_o = 0.3$ at this wavelength range). Considering the photon detection efficiency of the SSPDs as $\eta_d = 0.3$, we estimate the quantum efficiency of our SPE is roughly $\eta_q = I_\infty / (I_{total} \eta_c \eta_f \eta_o \eta_d) = 12\%$.

**Acknowledgments** We thank the discussion with Jurgen von Bardeleben. We acknowledge the support from the Singapore National Research Foundation through a Singapore 2015 NRF fellowship grant (NRF-NRFF2015-03) and its Competitive Research Program (CRP Award No. NRF-CRP14-2014-02), Singapore Ministry of Education (MOE2016-T2-2-077 and MOE2011-T3-1-005), A*Star QTE programme and a start-up grant (M4081441) from Nanyang Technological University. I. A. is the recipient of an Australian Research Council Discovery Early Career Research Award (Project Number DE130100592). Partial funding for this research was provided by the Air Force Office of Scientific Research, United States Air Force (grant FA2386-15-1-4044).

**Author Contributions** Y. Z., Z. W. built the optical set-up and performed the optical measurements. Y.Z., A.R., S. K. performed the data analysis. B. Z. and A.G. performed the wavelength modelling. Y.Z., I.A. and W. G designed the experiments and wrote the manuscript with contributions from all co-authors. All authors contributed to the discussion of the results.

**Author Information** The authors declare that they have no competing financial interests. Correspondence and requests for materials should be addressed to Igor.Aharonovich@uts.edu.au or wbgao@ntu.edu.sg.

**Figures and Figure Legends**

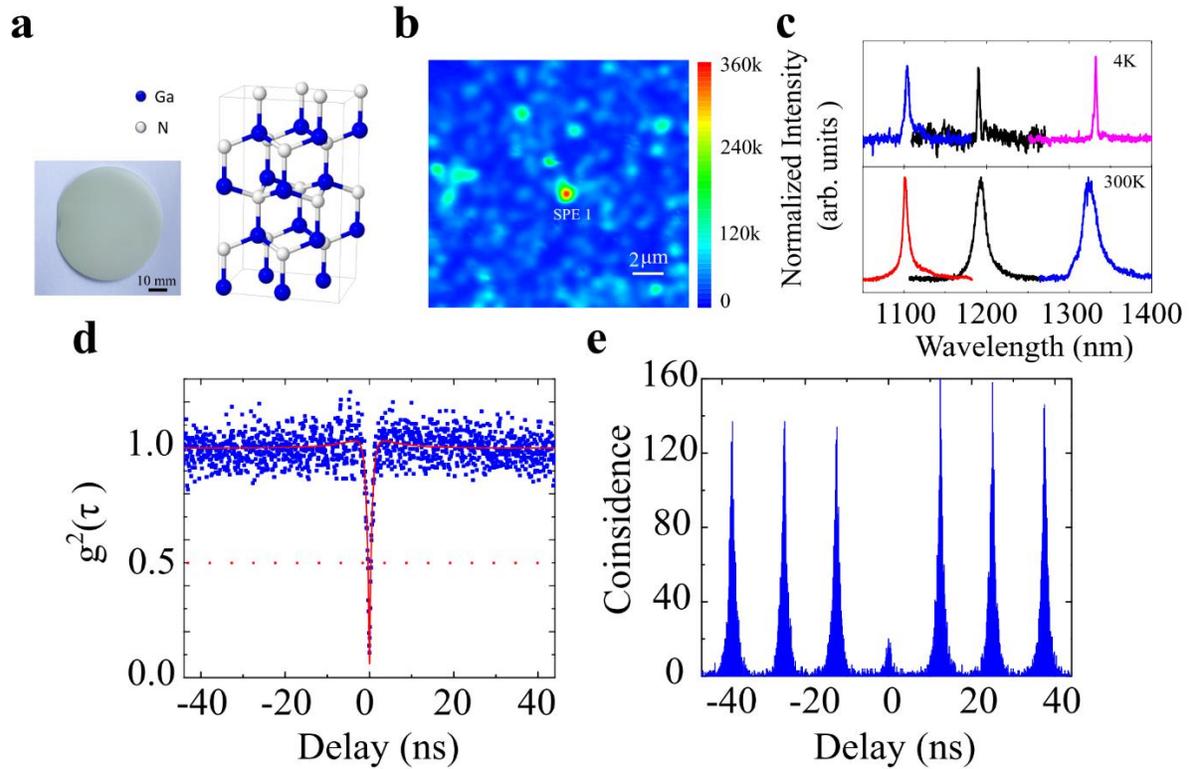

**Figure 1 | Infrared single photon emission in GaN. a,** Schematic illustration of gallium nitride crystal structure and an optical image of the GaN wafer. **b,** Confocal PL mapping with a single emitter SPE 1 in the centre of the map. **c**, Photoluminescence spectra of 6 infrared emitters, revealing the PL ranges from 1085 nm to 1340 nm. PL spectra from three emitters are taken at 4 K (upper panel) and 300 K (lower panel), respectively. Note that the emitters at 4K and RT are different. **d**, Second-order correlation measurement of the emission from SPE 1 under 950 nm cw laser excitation. The blue dots are the raw data without any background correction and the red curve is the fitting to a three level system, yielding $g^2(0) = 0.05 \pm 0.02$. **e**, Second-order correlation measurement of SPE 1 excited by pulsed laser with one picosecond pulse width and 80 MHz repetition rate, yielding $g^2(0) = 0.14 \pm 0.01$. The $g^2(\tau)$ measurements were recorded at room temperature.

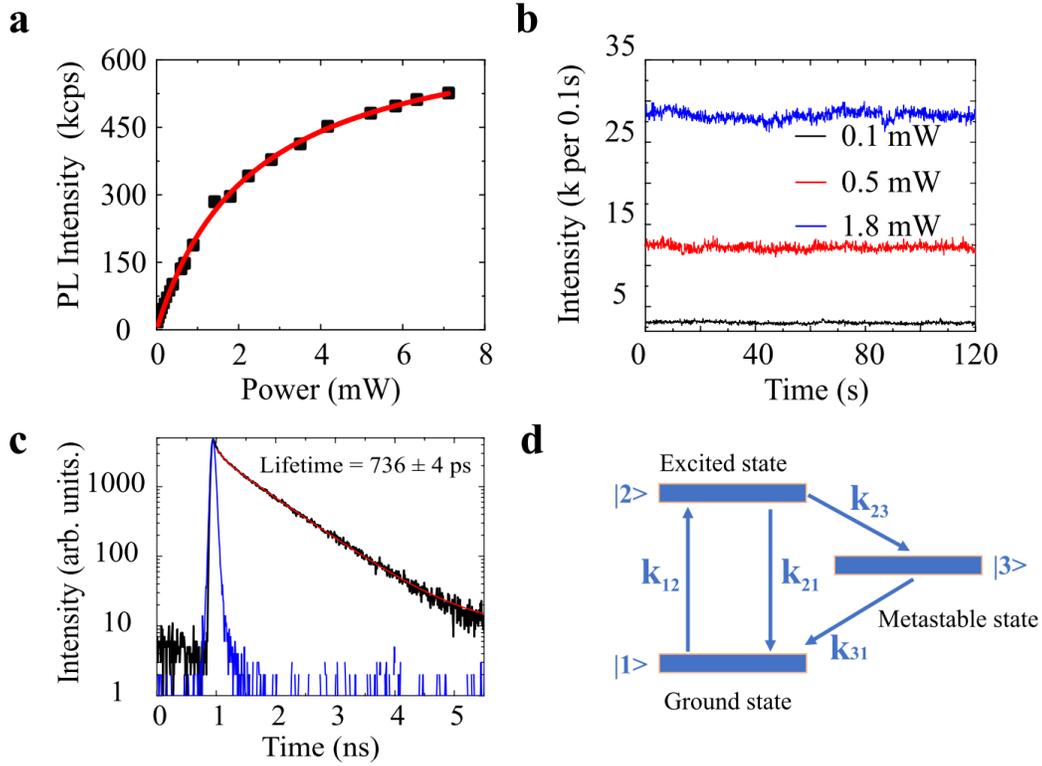

**Figure 2 | Optical properties of SPE 1. a**. Saturation curve of SPE 1 yielding a saturation power $P_s$ = 2.32 mW, and $I_\infty$ = 0.69M counts/s. **b**, photon-stability measurement at three different excitation powers of 0.1 mW (black curve), 0.8 mW (red curve) and 1.5 mW (blue curve), respectively, over a period of two minutes. The time resolution is 100 ms and no obvious blinking has been observed. **c**. fluorescence lifetime measurement of SPE1 (black curve) fit with a single exponent (red curve) yielding a lifetime $\tau_1$ = 736 $\pm$ 4 ps. The blue curve is the instrument response of the superconducting detector. **d**, A schematic diagram of a three level system, used to describe the emitter. Detailed analysis of the transition rate can be found in supplementary information.

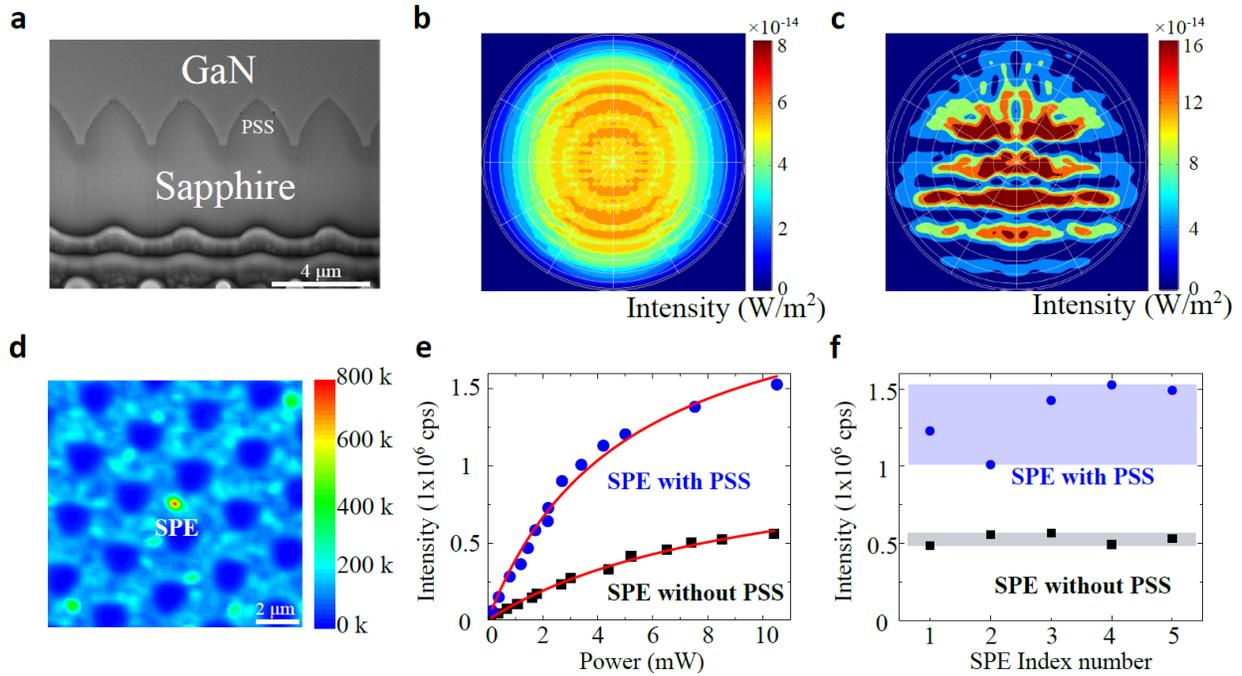

**Figure 3 | Enhancement of SPEs in GaN using a Patterned Sapphire Substrate. a,** SEM image of the cross section of the GaN grown on a PSS. Cone shapes of the PSS are clearly seen. **b,** Far-field radiation pattern of the in-plane dipole in the pristine GaN wafer. The circles represent collection half-angles from $10^0$ to $90^0$ (inner to outer circles). **c,** Far-field radiation pattern of the in-plane dipole in the GaN grown on top of a PSS. Note that the maximum intensity in c is two times higher than that in b. **d,** Confocal scan map of a SPE in GaN grown on PSS, with the blue circles corresponding to the PSS cones. **e,** Saturation curves comparing between a SPE in a pristine GaN (black squares) with a SPE embedded in a GaN grown on a PSS (blue circles). The saturated emission ($I_\infty$) for SPE with PSS reaches $2.33\times10^6$ counts/s compared to only $1.13\times10^6$ counts/s from pristine GaN. The red curve is the fitting function of the raw data. **f,** Comparison of count rate at 10 mw for five emitters from pristine GaN (black squares) and five emitters from GaN grown on the PSS (blue circles). The shaded rectangles are guide to the eye. The full saturation curves are presented in the supporting information.

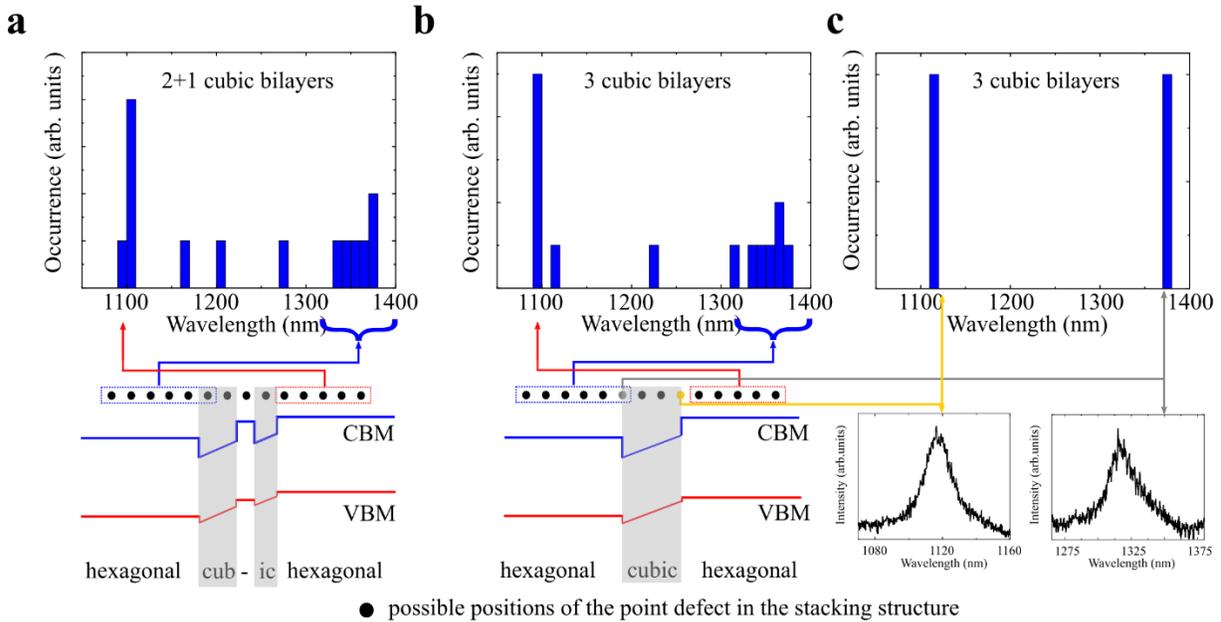

**Figure 4 | Numerical modelling of the SPEs. a, b,** Theoretical spectral distribution resulting from point defects distributed between cubic inclusions in a 2+1 bilayer configuration and a continuous 3 cubic bilayers, respectively. The blue columns correspond to the spectral position of the emitters' ZPL. The arrows connect possible positions of the point defect (distributed uniformly in each GaN bilayer in the neighbourhood of the cubic inclusion) and the resulting spectral positions of the ZPL. **c,** If the point defects are located solely at the interface, only two distinctive ZPLs are visible at ~1100 and 1350 nm. Our experimental data matches well scenario in **(a)** as well as a particular configuration in **(c)**. Two spectra are shown to exemplify the match between the predicted and the observed PL. CBM, VBM denotes the conduction band minimum and valence band maximum, respectively.

# Supplementary Information: Room-temperature solid state quantum emitters in the telecom range

## 1. Photoluminescence statistics

Room temperature photoluminescence (PL) of ten emitters is shown in Figure S1a. The ZPLs and linewidth of each emitter are summarized in Figure S1b.

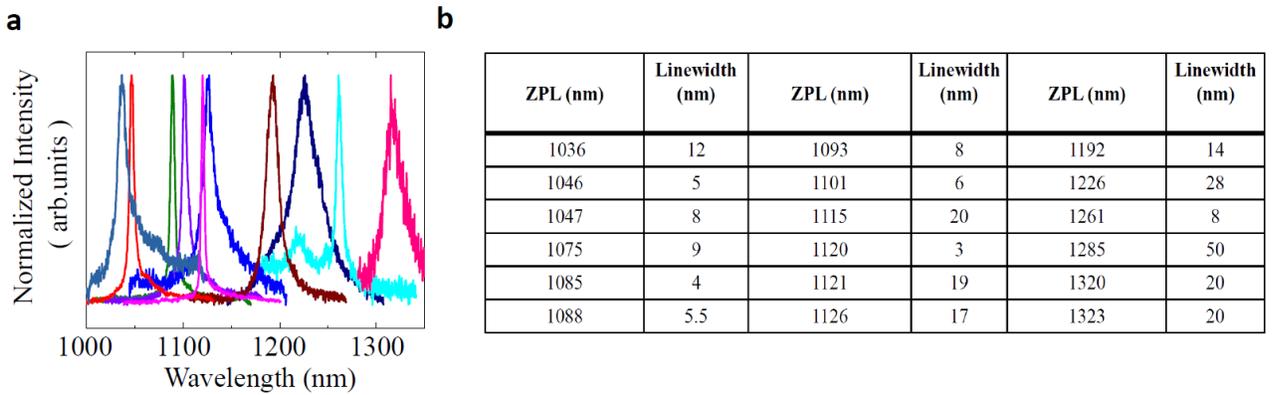

| ZPL (nm) | Linewidth (nm) | ZPL (nm) | Linewidth (nm) | ZPL (nm) | Linewidth (nm) |
|---|---|---|---|---|---|
| 1036 | 12 | 1093 | 8 | 1192 | 14 |
| 1046 | 5 | 1101 | 6 | 1226 | 28 |
| 1047 | 8 | 1115 | 20 | 1261 | 8 |
| 1075 | 9 | 1120 | 3 | 1285 | 50 |
| 1085 | 4 | 1121 | 19 | 1320 | 20 |
| 1088 | 5.5 | 1126 | 17 | 1323 | 20 |

**Figure S1. a,** Photoluminescence spectrum (PL) of 10 infrared emitters in GaN at room temperature. **b**, Summary of ZPL (zero phonon line) positions and linewidth of PL from 18 infrared emitters in GaN.

## 2. Second order correlation fitting procedure

To study energy level dynamics of SPE 1, we did a set of CW g2 measurements at 11 different excitation powers, four of them are shown in Figure S2a and the bunching effect is obvious in g2 curves. To explain this, at least three energy levels with a shelving state must be considered[1] (see Figure S2b). In Figure. S2b, $k_{12}$ is the excitation rate from energy level $|1\rangle$ to $|2\rangle$, which is proportional to the power ($k_{12} = \eta P$). Other values of $k_{xy}$ represent the decay rate from energy level $|x\rangle$ to $|y\rangle$. The three level system will result in second order correlation function as

$$g^2(\tau) = 1-(1+\beta)*exp(-|\tau|/\tau 1)+ \beta*exp(-|\tau|/\tau 2), \quad (1)$$

Here $\beta$, $\tau_1$ and $\tau_2$ are

$$\beta = \frac{1-\tau_2 k_{31}}{k_{31}(\tau_2-\tau_1)}, \quad \tau_{1,2} = \frac{2}{A \pm \sqrt{A^2-4B}}, \quad (2)$$

where $A = k_{12}+k_{21}+k_{23}+k_{31}$, $B = k_{12}(k_{23}+k_{31})+k_{31}(k_{21}+k_{23})$.

Experimentally, in order to get the transition rate $k_{xy}$, we first get the set of experimental values of $\beta$, $\tau_1$ and $\tau_2$ by fitting the data in figure. S2a with equation (1). The values are shown in Fig. S2c, S2d, S2e. We then proceed to obtain the values of $k_{xy}$. First, $k_{31}$ can be directly calculated with $k_{31} = \frac{1}{[\beta(\tau_2-\tau_1)+\tau_2]}$ derived from equation (2). Second, $\tau_1$ and $\tau_2$ as a function of power are fitted with $k_{23}$, $k_{21}$ and $\eta$ are as free parameters.

The fitting matches with the experimental data well, which shows the correctness of the model. Based on the fitting, the excited state lifetime $(k_{21}+k_{23})^{-1}$ is around $776 \pm 39$ ps which roughly matches the measured lifetime value $736 \pm 4$ ps with a pulse laser.

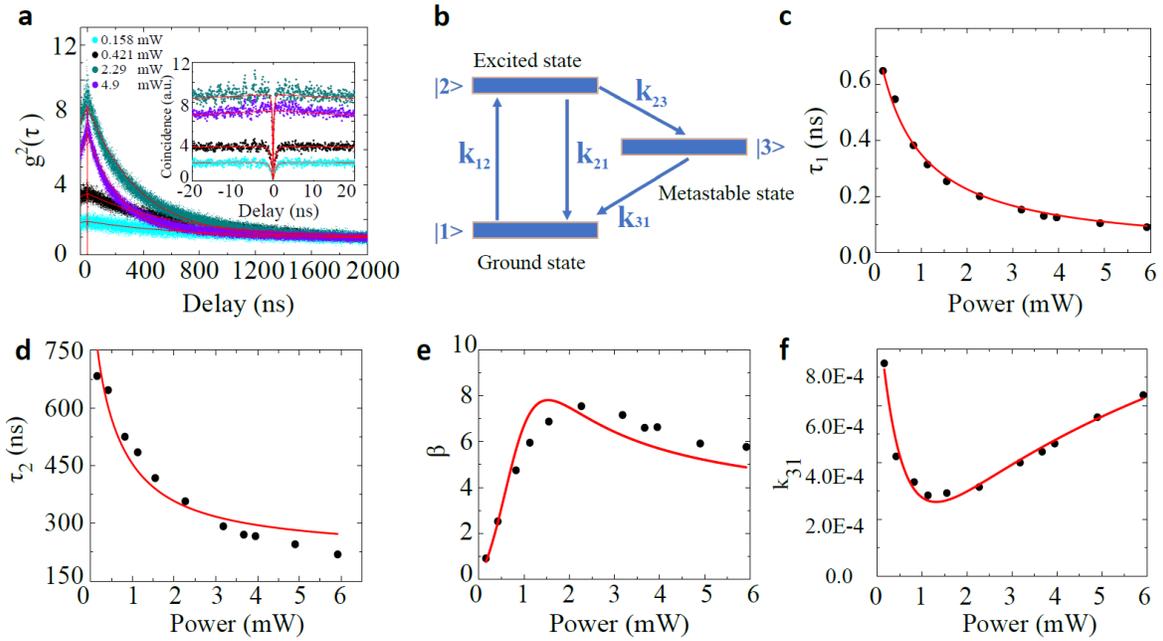

**Figure S2. Photon antibunching measurement of SPE 1**. **a**, CW $g^{(2)}(\tau)$ measurement at 4 different excitation powers, the red solid curve is the fitting curve. Inset figure is the zoom-in of $g^{(2)}(\tau)$ between -20 ns to 20 ns. **b**, Schematic of three energy level diagram for the emitter. A metastable state $|3\rangle$ has been included. $k_{xy}$ represent the decay rate from energy level $|x\rangle$ to . **c-f,** Fitting parameters $\tau_1$, $\tau_2$, $\beta$, and $k_{31}$ as a function of the excitation power. The red curve is the fitting based on our models described above and $k_{31}$ fits with a combination of exponential decay and linear function as $\dfrac{a}{Exp(bx)} + \dfrac{dx}{(x+c)}$.

## 3. Model to compare SPEs with and without PSS

We built a numerical model to estimate the enhancement by the PSS structure. Here we provide some complementary information of the model. As shown in Figure S3a, we measured the emission polarization of SPE 1. The emission polarization is measured by rotating half-waveplate before the polarizer in the collection arm of the confocal set up, while fixing the excitation polarizations. It shows that the emitter has a single polarization comment and therefore could be treated as a dipole in the model. Figure S3b, S3c are the top view and side view of the model we created respectively in the Lumerical simulation.

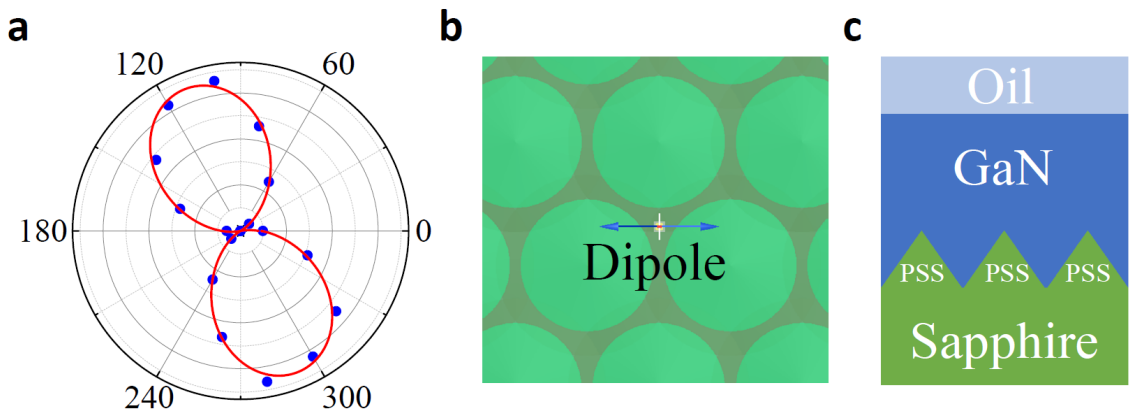

**Figure S3. Model of SPE in GaN with PSS structure.** a, Polarization measurement of the emission from SPE 1 in the main text. The angles represent emission polarization angles, showing a single polarization component for the emitter. The red curve are the fitting curve with

$y = y_0 + A\cos^2(ax+\varphi)$. b, Vertical view of the model built, an electric dipole is sitting in the middle of three PSS cones. c, Side view of the model built, the dimension are based on real geometry and SEM images. The dipole is located at 500 nm above Sapphire-GaN interface.

## 4. Comparison between SPEs in pristine GaN and SPEs in GaN grown on Patterned Sapphire Substrates (PSS)

To give a detailed comparison between SPEs in pristine GaN and in GaN grown on PSS, first we found find 5 SPEs in pristine GaN. As shown in Fig.S3a, ZPLs of these 5 SPEs are around 1100 nm and Fig. S4b shows that at 10.5 mW, the detected photon counts is around 0.5 M cps. Figure S5 shows $g^2(0)$ recorded from these five different SPEs in pristine GaN. All $g^2(0)$ is well below 0.5, confirm that they are indeed single emitters in pristine GaN.

As a comparison, we intentionally find SPEs in GaN grown on PSS with similar spectra wavelength (as shown in Fig. S6a). From the saturation curve in Fig. S6b, at 10.5 mW, all 5 SPEs reach more than $1\times10^6$ counts/s, shows an obvious enhancement. Second order correlation measurement in Figure S7 confirms all 5 emitters we have found are SPEs. We also measured the lifetimes of all these SPEs both in pristine GaN and in GaN grown on PSS. They are summarized in Table S1. No big difference has been observed in terms of lifetime. Also, $I_\infty$ from the saturation curve fitting in Fig. S4b and Fig. S6b are included.

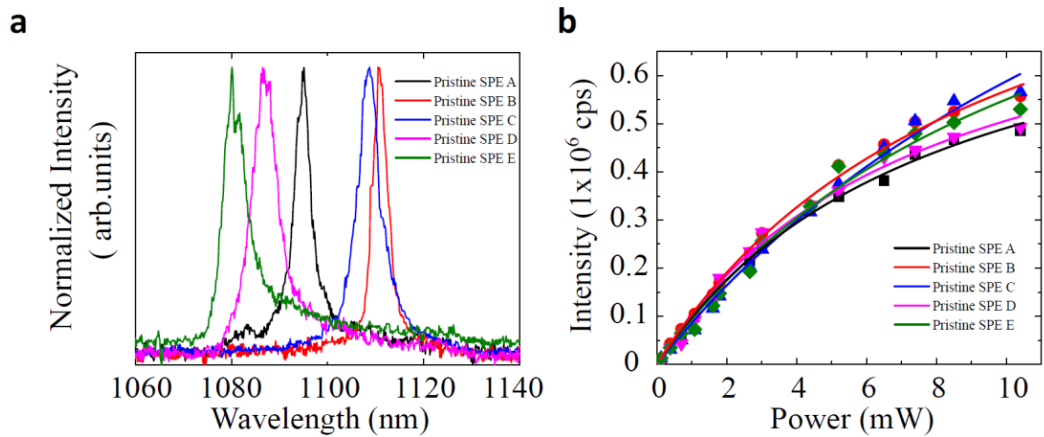

**Figure S4. a,** PL of 5 SPEs in pristine GaN. **b,** Saturation curves for these SPEs. For each saturation curve, the data is fitted well with an equation $I(P)=I_\infty \times P/(P+P_s)$, where $I(P)$ is the intensity count rate, $P$ is the excitation power; $P_s$ is the excitation power at saturation and $I_\infty$ is the emitter maximum counts.

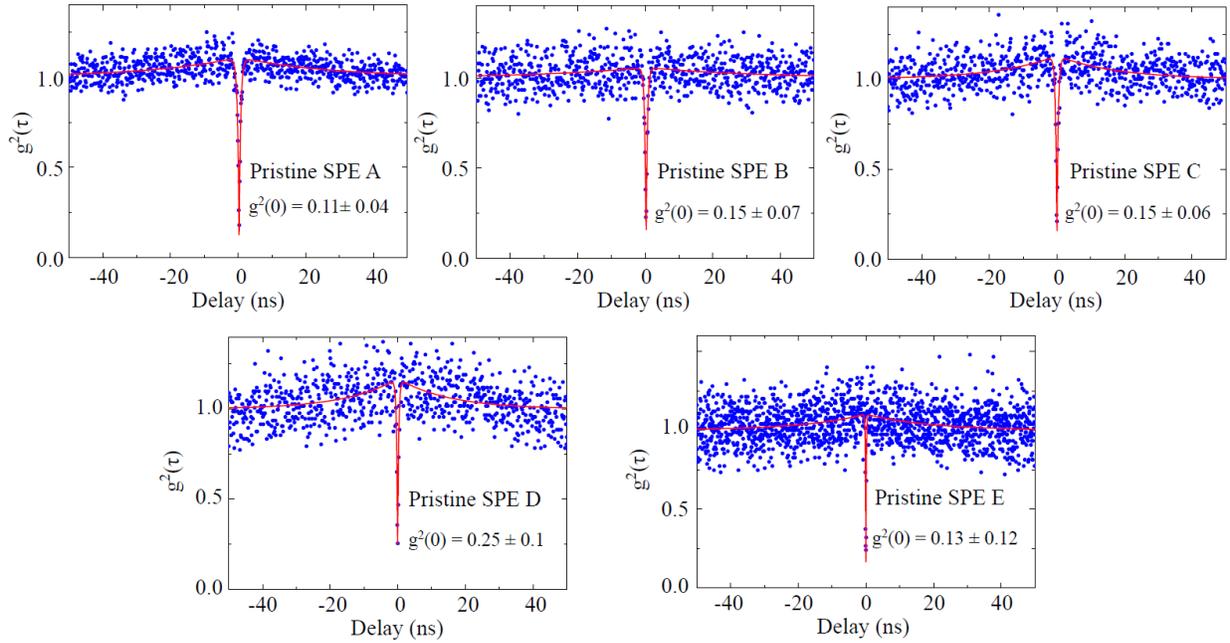

**Figure S5.** Second order autocorrelation function of 5 SPEs in pristine GaN. The red curve is the fitting curves with equation (1) in the main text. The g² (0) value is given in each graph.

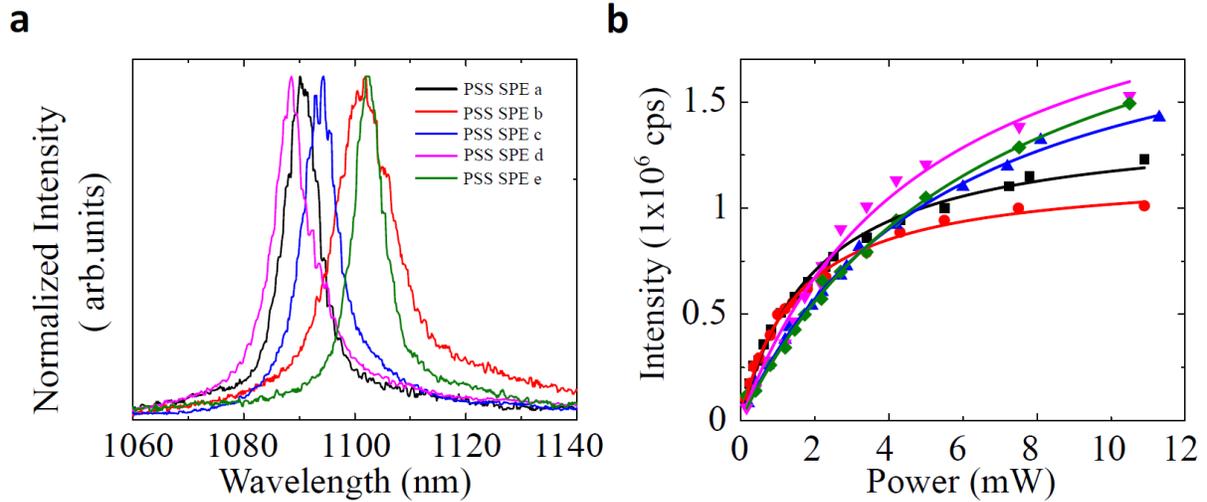

**Figure S6. a,** PL of 5 SPEs in GaN grown on PSS. **b,** Saturation curves for these SPEs in GaN on PSS. Fitting curves use the same equation as in Figure S4. The same color corresponds to the same emitter as in **a**.

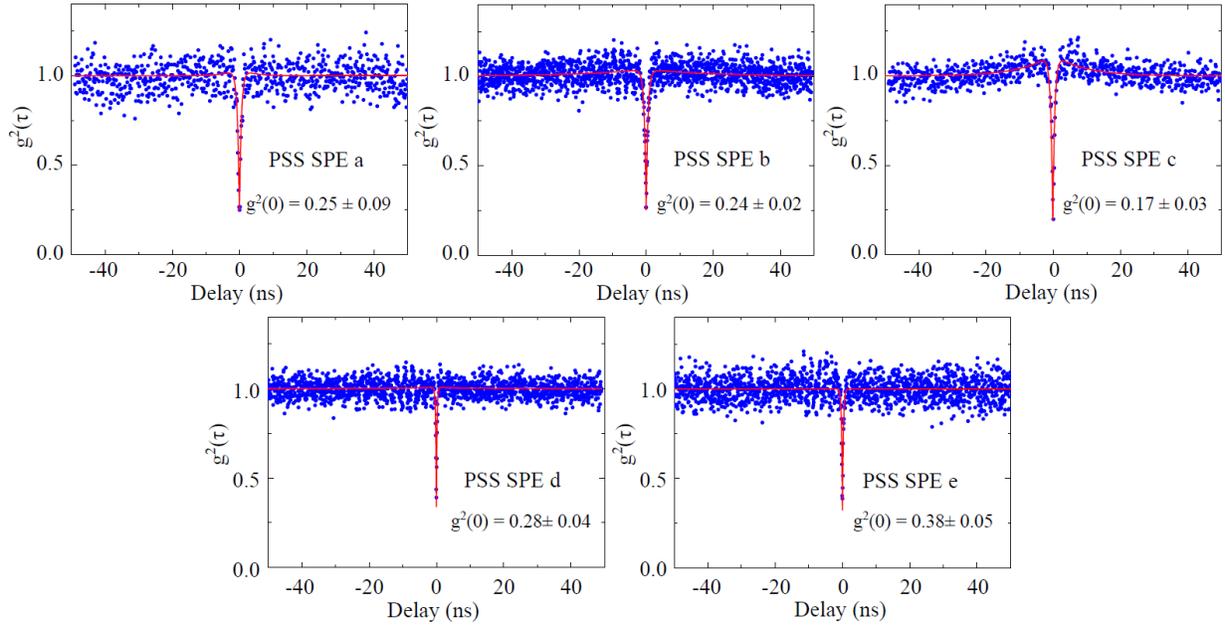

**Figure S7.** Second order autocorrelation function of 5 SPEs in GaN grown on PSS. The red curve is the fitting with a three level model. The $g^2(0)$ value is given in each graph. All the values are below 0.5, proving their SPE character.

| SPEs in pristine GaN | Lifetime (ps) | $I_\infty$ from saturation curve fitting | SPEs in GaN grown on PSS | Lifetime (ps) | $I_\infty$ from saturation curve fitting |
|---|---|---|---|---|---|
| A | 484 | 0.89 M | a | 482 | 1.4 M |
| B | 448 | 1.13 M | b | 442 | 1.17 M |
| C | 348 | 1.67 M | c | 405 | 2.13 M |
| D | 456 | 0.89 M | d | 348 | 2.33 M |
| E | 265 | 1.19 M | e | 396 | 2.45 M |

**Table S1.** Summary of measured lifetimes and maximum counts of SPE A-E in pristine GaN and SPE a-e in GaN grown on PSS. The maximum counts are obtained from the fitting of the curves

in Fig. S4b and Fig. S6b. The tables show that the lifetimes in two samples are similar while the counts for PSS sample is larger than pristine sample in average.

## 5. Modelling of a proposed defect structure

The PL energies are calculated with a simple quasi one-dimension model, where a point defect resides in the neighbourhood of a cubic inclusion of three bilayers in wurzite GaN. Photoluminescence is the result of an exciton recombination, whose hole part is tightly localized at the point defect while the loosely localized electron's position is more heavily influenced by the potential lineup generated by the stacking sequence of cubic and hexagonal GaN bilayers. Thus, the binding energy of the exciton and PL wavelength are determined by the relative position of the point defect with respect to the cubic inclusion.

The energetics of the exciton is a result of the interplay of several partial effects. The original ZPL energy which would be emitted by the point defect in a homogeneous (i.e. fully wurtzite of fully cubic) environment is modified by the varying Coulomb energy when the distance between the respective positions of the hole (localized at the point defect) and the electron (diffusely centered at the low end of the triangular potential of the cubic part) itself varies. The energy is also a function of valence band maximum (VBM) and conduction band minimum (CBM) energy levels at the locations of exciton parts, and in second order, it is also a function of the shape of the electron wavefunction, which results from the CBM profile and the hole's Coulomb potential.

As can be seen, a combination of a point defect of originally 1350 nm photoluminescence and three bilayers of cubic inclusion show the characteristics of measured spectra. The nature of the point defect is responsible roughly for the longest wavelength at which a ZPL line may appear;

whereas the width of the cubic inclusion adjusts the interval spanned by ZPL wavelengths, when photoluminescence comes from the point defects in varying positions (equally distributed in the cubic part, as well as around it).

The material properties of GaN, as well as relative positions of band gaps and most notably the 2.9 MV/cm change of electrical field at the hexagonal-cubic boundary, due to the charge density arising there, have been taken from Ref. 2 (electric field) and Ref. 3 (material properties).